\documentclass[conference]{IEEEtran}
\IEEEoverridecommandlockouts
\usepackage{cite}
\usepackage{amsmath,amssymb,amsfonts}
\usepackage{algorithmic}
\usepackage{graphicx}
\usepackage{textcomp}
\usepackage{xcolor}
\usepackage{tikz}
\def\BibTeX{{\rm B\kern-.05em{\sc i\kern-.025em b}\kern-.08em
    T\kern-.1667em\lower.7ex\hbox{E}\kern-.125emX}}
\begin{document}
\title{Grid-forming Control of Converter Infinite Bus System: Modeling by Data-driven Methods 
\thanks{This material is based upon work supported by the National Science Foundation under Grant Number 2328241. Any opinions, findings, and conclusions or recommendations expressed in this material are those of the authors and do not necessarily reflect the views of the National Science Foundation.}
\thanks{A.B. Javadi, and P. Pong are with the Helen and John C. Hartmann Department of Electrical and Computer Engineering, New Jersey Institute of Technology, Newark, NJ 07102 USA (e-mail: \{aj772, philip.pong\}@njit.edu).}
}

\author{
    \IEEEauthorblockN{Amir Bahador Javadi, \IEEEmembership{Student Member, IEEE,} and Philip Pong,~\IEEEmembership{Senior~Member,~IEEE}
    }
       }

\maketitle

\begin{abstract}
This study explores data-driven modeling techniques to capture the dynamics of a grid-forming converter-based infinite bus system, critical for renewable-integrated power grids. Using sparse identification of nonlinear dynamics and deep symbolic regression, models were generated from synthetic data simulating key disturbances in active power, reactive power, and voltage references. Deep symbolic regression demonstrated more accuracy in capturing complex system dynamics, though it required substantially more computational time than sparse identification of nonlinear dynamics. These findings suggest that while deep symbolic regression offers high fidelity, sparse identification of nonlinear dynamics provides a more computationally efficient approach, balancing accuracy and runtime for real-time grid applications.
\end{abstract}


\section{Introduction}
The transition to renewable energy sources has necessitated significant shifts in power systems, particularly in how grid stability and reliability are maintained. One of the pivotal technologies in this transformation is the grid-forming converter, a type of inverter-based resource essential for supporting power systems dominated by renewable generation. Unlike traditional synchronous generators, grid-forming converters enable power systems to operate in a decentralized and resilient manner, even with high penetration of renewables. However, the modeling and control of grid-forming converters are challenging, especially under dynamic and fault conditions, due to the complex nonlinearities and interdependencies present in their operation \cite{b1}.

Grid-forming converters differ fundamentally from grid-following converters in terms of their control objectives and operational characteristics. While grid-following converters are designed to synchronize to an existing grid voltage and frequency, relying on a stable grid to function, grid-forming converters can independently establish and regulate the grid’s frequency and voltage. This capability allows grid-forming converters to maintain stability in weak or islanded grid conditions where synchronous machines may no longer be present or feasible. The distinction between grid-forming and grid-following converters is crucial as the grid increasingly depends on inverter-based resources for maintaining stability \cite{b2}.

Despite the benefits, the modeling and control of grid-forming converters pose significant challenges. The nonlinear dynamics, interactions with other grid components, and the need for robust stability under various loading conditions require sophisticated models that can capture these behaviors with high fidelity. Traditional modeling approaches often fall short in representing the nuanced dynamics of grid-forming converters, especially as power grids become increasingly complex. Furthermore, grid-forming converters must operate reliably under diverse conditions, including high renewable penetration, variable loads, and fault disturbances, making real-time adaptive modeling both a necessity and a challenge \cite{b3}.

To address these challenges, data-driven modeling techniques, such as sparse identification of nonlinear dynamics (SINDy) and deep symbolic regression (DSR), offer promising solutions. SINDy is particularly useful in identifying governing equations from time-series data, providing interpretable models that capture essential system dynamics while maintaining computational efficiency. This approach is advantageous for grid-forming converters, where capturing transient and nonlinear behaviors is crucial \cite{b4}. DSR, on the other hand, leverages neural networks and genetic programming to discover mathematical expressions that best describe system dynamics. This method is well-suited for exploring the complex interactions and unknown dependencies present in grid-forming converters, offering a more flexible framework for model discovery \cite{b5}.

In this study, the SINDy and DSR frameworks were applied to a modified single machine infinite bus system, replacing the conventional generator with a converter-based resource. Simulations under a grid-forming control strategy were conducted to generate synthetic data and evaluate the effectiveness of SINDy and DSR in accurately capturing the system's dynamics. Our findings indicate that DSR yields a robust, interpretable model of the system, offering valuable insights into the behavior of converter-based systems operating in grid-forming control mode. This work advances efforts to develop sophisticated modeling tools that enhance the stability and reliability of future power grids.

\section{Converter Infinite Bus System}
\label{sec2}
The single machine infinite bus system is a foundational concept in power system analysis. In this study, the conventional synchronous generator was replaced with a converter-based resource, maintaining the essential characteristics of the traditional system. The converter-based resource interfaces with an infinite bus, which represents a large, stable power grid characterized by constant voltage and frequency. This connection is facilitated through a lossless transmission line and the resource is equipped with an LCL filter to reduce harmonic distortion and enhance the power quality of the output voltage \cite{b6}.

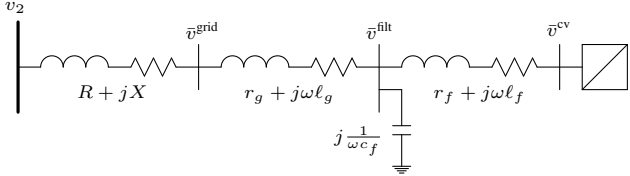
\begin{figure}[htbp]
\centering
\begin{tikzpicture}[line cap=round,line join=round,=>triangle 45,x=1cm,y=1cm,scale=3]
\clip(3.6,3.4) rectangle (6.6,4.4);
\draw (3.70,4.34) node[anchor=north west] {\scriptsize $\bar{v}_{2}$};
\draw (4.50,4.24) node[anchor=north west] {\scriptsize $\bar{v}^{\textnormal{grid}}$};
\draw (5.30,4.24) node[anchor=north west] {\scriptsize $\bar{v}^{\textnormal{filt}}$};
\draw (6.10,4.24) node[anchor=north west] {\scriptsize $\bar{v}^{\textnormal{cv}}$};
\draw (4.02,3.96) node[anchor=north west] {\scriptsize $R+jX$};
\draw (4.76,3.96) node[anchor=north west] {\scriptsize $r_{g}+j\omega\ell_{g}$};
\draw (5.60,3.96) node[anchor=north west] {\scriptsize $r_{f}+j\omega\ell_{f}$};
\draw (5.16,3.78) node[anchor=north west] {\scriptsize $j\frac{1}{\omega c_{f}}$};
\draw [shift={(3.95,4)},line width=0.4pt]  plot[domain=0:3.141592653589793,variable=\t]({1*0.05*cos(\t r)+0*0.05*sin(\t r)},{0*0.05*cos(\t r)+1*0.05*sin(\t r)});
\draw [shift={(4.05,4)},line width=0.4pt]  plot[domain=0:3.141592653589793,variable=\t]({1*0.05*cos(\t r)+0*0.05*sin(\t r)},{0*0.05*cos(\t r)+1*0.05*sin(\t r)});
\draw [shift={(4.15,4)},line width=0.4pt]  plot[domain=0:3.141592653589793,variable=\t]({1*0.05*cos(\t r)+0*0.05*sin(\t r)},{0*0.05*cos(\t r)+1*0.05*sin(\t r)});
\draw [line width=0.4pt] (4.2,4)-- (4.3,4);
\draw [line width=0.4pt] (4.3,4)-- (4.32,4.04);
\draw [line width=0.4pt] (4.32,4.04)-- (4.36,3.96);
\draw [line width=0.4pt] (4.36,3.96)-- (4.4,4.04);
\draw [line width=0.4pt] (4.4,4.04)-- (4.44,3.96);
\draw [line width=0.4pt] (4.44,3.96)-- (4.48,4.04);
\draw [line width=0.4pt] (4.48,4.04)-- (4.5,4);
\draw [line width=0.4pt] (4.5,4)-- (4.7,4);
\draw [line width=0.4pt] (4.6,4.1)-- (4.6,3.9);
\draw [line width=1.2pt] (3.8,4.2)-- (3.8,3.8);
\draw [line width=0.4pt] (3.9,4)-- (3.8,4);
\draw [shift={(4.75,4)},line width=0.4pt]  plot[domain=0:3.141592653589793,variable=\t]({1*0.05*cos(\t r)+0*0.05*sin(\t r)},{0*0.05*cos(\t r)+1*0.05*sin(\t r)});
\draw [shift={(4.85,4)},line width=0.4pt]  plot[domain=0:3.141592653589793,variable=\t]({1*0.05*cos(\t r)+0*0.05*sin(\t r)},{0*0.05*cos(\t r)+1*0.05*sin(\t r)});
\draw [shift={(4.95,4)},line width=0.4pt]  plot[domain=0:3.141592653589793,variable=\t]({1*0.05*cos(\t r)+0*0.05*sin(\t r)},{0*0.05*cos(\t r)+1*0.05*sin(\t r)});
\draw [line width=0.4pt] (5.1,4)-- (5.12,4.04);
\draw [line width=0.4pt] (5.12,4.04)-- (5.16,3.96);
\draw [line width=0.4pt] (5.16,3.96)-- (5.2,4.04);
\draw [line width=0.4pt] (5.2,4.04)-- (5.24,3.96);
\draw [line width=0.4pt] (5.24,3.96)-- (5.28,4.04);
\draw [line width=0.4pt] (5.28,4.04)-- (5.3,4);
\draw [line width=0.4pt] (5,4)-- (5.1,4);
\draw [line width=0.4pt] (5.3,4)-- (5.5,4);
\draw [line width=0.4pt] (5.4,4.1)-- (5.4,3.8);
\draw [line width=0.4pt] (5.5,3.9)-- (5.4,3.9);
\draw [line width=0.4pt] (5.5,3.9)-- (5.5,3.74);
\draw [shift={(5.55,4)},line width=0.4pt]  plot[domain=0:3.141592653589793,variable=\t]({1*0.05*cos(\t r)+0*0.05*sin(\t r)},{0*0.05*cos(\t r)+1*0.05*sin(\t r)});
\draw [shift={(5.65,4)},line width=0.4pt]  plot[domain=0:3.141592653589793,variable=\t]({1*0.05*cos(\t r)+0*0.05*sin(\t r)},{0*0.05*cos(\t r)+1*0.05*sin(\t r)});
\draw [shift={(5.75,4)},line width=0.4pt]  plot[domain=0:3.141592653589793,variable=\t]({1*0.05*cos(\t r)+0*0.05*sin(\t r)},{0*0.05*cos(\t r)+1*0.05*sin(\t r)});
\draw [line width=0.4pt] (5.8,4)-- (5.9,4);
\draw [line width=0.4pt] (5.9,4)-- (5.92,4.04);
\draw [line width=0.4pt] (5.92,4.04)-- (5.96,3.96);
\draw [line width=0.4pt] (5.96,3.96)-- (6,4.04);
\draw [line width=0.4pt] (6,4.04)-- (6.04,3.96);
\draw [line width=0.4pt] (6.04,3.96)-- (6.08,4.04);
\draw [line width=0.4pt] (6.08,4.04)-- (6.1,4);
\draw [line width=0.4pt] (6.1,4)-- (6.3,4);
\draw [line width=0.4pt] (6.2,4.1)-- (6.2,3.9);
\draw [line width=0.4pt] (6.3,4.1)-- (6.3,3.9);
\draw [line width=0.4pt] (6.3,3.9)-- (6.5,3.9);
\draw [line width=0.4pt] (6.5,3.9)-- (6.5,4.1);
\draw [line width=0.4pt] (6.5,4.1)-- (6.3,4.1);
\draw [line width=0.4pt] (6.3,3.9)-- (6.5,4.1);
\draw [line width=0.4pt] (5.46,3.74)-- (5.54,3.74);
\draw [line width=0.4pt] (5.46,3.7)-- (5.54,3.7);
\draw [line width=0.4pt] (5.5,3.56)-- (5.5,3.7);
\draw [line width=0.4pt] (5.46,3.56)-- (5.54,3.56);
\draw [line width=0.4pt] (5.47,3.55)-- (5.53,3.55);
\draw [line width=0.4pt] (5.48,3.54)-- (5.52,3.54);
\draw [line width=0.4pt] (5.49,3.53)-- (5.51,3.53);
\end{tikzpicture}
\caption{One-line diagram of the converter infinite bus system under grid-forming control mode.}
\label{CIB}
\end{figure}

The one-line schematic of the converter infinite bus system illustrated in Fig. \ref{CIB}. In this diagram, $\bar{v}_{2}$, $\bar{v}^{\textnormal{grid}}$, $\bar{v}^{\textnormal{filt}}$, and $\bar{v}^{\textnormal{cv}}$ denote the voltage at the infinite bus, the output terminal voltage of the LCL filter connected to the transmission line (or point of common coupling), the LCL filter voltage, and the voltage at the output of the converter, respectively. The parameters $r_{g}$ and $\ell_{g}$ represent the resistance and reactance of the LCL filter on the grid side, while $r_{f}$ and $\ell_{f}$ denote the resistance and reactance on the converter side of the filter. Additionally, $c_{f}$ refers to the capacitance of the LCL filter. The power flow between the converter-based resource and the infinite bus system, along with the dynamics of the LCL filter, are governed by Eqs. \eqref{1}, and \eqref{2}.
\begin{align}
\textnormal{Power grid}
&\begin{cases}
\bar{v}_{2} &=v_{2}e^{j\theta_{2}} \\
\bar{v}^{\textnormal{grid}} &=v^{\textnormal{grid}}e^{j\theta^{\textnormal{grid}}} = v_{r}^{\textnormal{grid}} + j v_{i}^{\textnormal{grid}} \\
\bar{v}^{\textnormal{filt}} &=v^{\textnormal{filt}}e^{j\theta^{\textnormal{filt}}} = v_{r}^{\textnormal{filt}} + j v_{i}^{\textnormal{filt}} \\
\bar{v}^{\textnormal{cv}} &=v^{\textnormal{cv}}e^{j\theta^{\textnormal{cv}}} = v_{r}^{\textnormal{cv}} + j v_{i}^{\textnormal{cv}} \\
0&= \left(v^{\textnormal{grid}}\right)^{2}G_{\textnormal{11}} + v^{\textnormal{grid}}v_{2}G_{\textnormal{12}}\cos{\left(\theta^{\textnormal{grid}}-\theta_{2}\right)} \\&+ v^{\textnormal{grid}}v_{2}B_{\textnormal{12}}\sin{\left(\theta^{\textnormal{grid}}-\theta_{2}\right)} +\left(v^{\textnormal{grid}}\right)^{2}G_{\textnormal{ff}} \\&+ v^{\textnormal{grid}}v^{\textnormal{filt}}G_{\textnormal{1f}}\cos{\left(\theta^{\textnormal{grid}}-\theta^{\textnormal{filt}}\right)} \\&+ v^{\textnormal{grid}}v^{\textnormal{filt}}B_{\textnormal{1f}}\sin{\left(\theta^{\textnormal{grid}}-\theta^{\textnormal{filt}}\right)}\\
Y_{\textnormal{1}} &= G_{\textnormal{11}} + jB_{\textnormal{11}} = \frac{1}{R+jX} \\
G_{\textnormal{12}} &= -G_{\textnormal{11}}\\
B_{\textnormal{12}} &= -B_{\textnormal{11}} \\
Y_{\textnormal{f}} &= G_{\textnormal{ff}} + jB_{\textnormal{ff}} = \frac{1}{r_{g}+j\omega\ell_{g}} \\
G_{\textnormal{1f}} &= -G_{\textnormal{ff}}\\
B_{\textnormal{1f}} &= -B_{\textnormal{ff}}
\label{1}
\end{cases}
\end{align}
\begin{align}
\textnormal{LCL Filter}
&\begin{cases}
\frac{\ell_{f}}{\omega_{b}} \frac{d}{dt}{i}_{r}^{\textnormal{cv}}&= v_{r}^{\textnormal{cv}} - v_{r}^{\textnormal{filt}} - r_{f}i_{r}^{\textnormal{cv}} + \omega_{s}\ell_{f}i_{i}^{\textnormal{cv}} \\
\frac{\ell_{f}}{\omega_{b}} \frac{d}{dt}{i}_{i}^{\textnormal{cv}}&= v_{i}^{\textnormal{cv}} - v_{i}^{\textnormal{filt}} - r_{f}i_{i}^{\textnormal{cv}} - \omega_{s}\ell_{f}i_{r}^{\textnormal{cv}} \\
\frac{c_{f}}{\omega_{b}} \frac{d}{dt}{v}_{r}^{\textnormal{filt}}&= i_{r}^{\textnormal{cv}} - i_{r}^{\textnormal{filt}} + \omega_{s}c_{f}v_{i}^{\textnormal{filt}} \\
\frac{c_{f}}{\omega_{b}} \frac{d}{dt}{v}_{i}^{\textnormal{filt}}&= i_{i}^{\textnormal{cv}} - i_{i}^{\textnormal{filt}} - \omega_{s}c_{f}v_{r}^{\textnormal{filt}} \\
\frac{\ell_{g}}{\omega_{b}} \frac{d}{dt}{i}_{r}^{\textnormal{filt}}&= v_{r}^{\textnormal{filt}} - v_{r}^{\textnormal{grid}} - r_{g}i_{r}^{\textnormal{filt}} + \omega_{s}\ell_{g}i_{i}^{\textnormal{filt}} \\
\frac{\ell_{g}}{\omega_{b}} \frac{d}{dt}{i}_{i}^{\textnormal{filt}}&= v_{i}^{\textnormal{filt}} - v_{i}^{\textnormal{grid}} - r_{g}i_{i}^{\textnormal{filt}} - \omega_{s}\ell_{g}i_{r}^{\textnormal{filt}} \\
v_{d}^{\textnormal{filt}}&= \sin{(\theta^{\textnormal{oc}}}+\frac{\pi}{2})v_{r}^{\textnormal{filt}} - \cos{(\theta^{\textnormal{oc}}}+\frac{\pi}{2})v_{i}^{\textnormal{filt}} \\
v_{q}^{\textnormal{filt}}&= \cos{(\theta^{\textnormal{oc}}}+\frac{\pi}{2})v_{r}^{\textnormal{filt}} + \sin{(\theta^{\textnormal{oc}}}+\frac{\pi}{2})v_{i}^{\textnormal{filt}}\\
i_{d}^{\textnormal{filt}}&= \sin{(\theta^{\textnormal{oc}}}+\frac{\pi}{2})i_{r}^{\textnormal{filt}} - \cos{(\theta^{\textnormal{oc}}}+\frac{\pi}{2})i_{i}^{\textnormal{filt}} \\
i_{q}^{\textnormal{filt}}&= \cos{(\theta^{\textnormal{oc}}}+\frac{\pi}{2})i_{r}^{\textnormal{filt}} + \sin{(\theta^{\textnormal{oc}}}+\frac{\pi}{2})i_{i}^{\textnormal{filt}} \\
i_{d}^{\textnormal{cv}}&= \sin{(\theta^{\textnormal{oc}}}+\frac{\pi}{2})i_{r}^{\textnormal{cv}} - \cos{(\theta^{\textnormal{oc}}}+\frac{\pi}{2})i_{i}^{\textnormal{cv}} \\
i_{q}^{\textnormal{cv}}&= \cos{(\theta^{\textnormal{oc}}}+\frac{\pi}{2})i_{r}^{\textnormal{cv}} + \sin{(\theta^{\textnormal{oc}}}+\frac{\pi}{2})i_{i}^{\textnormal{cv}}
\label{2}
\end{cases}
\end{align}
Moreover, the dynamics of the converter are represented by the average model, as shown in Eq. \eqref{eq3}.
\begin{align}
\textnormal{Converter}
&\begin{cases}
v_{r}^{\textnormal{cv}} &= \sin{(\theta^{\textnormal{oc}}}+\frac{\pi}{2})v_{d}^{\textnormal{cv,ref}} + \cos{(\theta^{\textnormal{oc}}}+\frac{\pi}{2})v_{q}^{\textnormal{cv,ref}} 
\\
v_{i}^{\textnormal{cv}} &= -\cos{(\theta^{\textnormal{oc}}}+\frac{\pi}{2})v_{d}^{\textnormal{cv,ref}} + \sin{(\theta^{\textnormal{oc}}}+\frac{\pi}{2})v_{q}^{\textnormal{cv,ref}}
\label{eq3}
\end{cases}
\end{align}

Grid-forming control mode is a sophisticated control strategy where the converter takes an active role in regulating both voltage and frequency, effectively creating a stable reference for the grid. This control mode establishes the grid parameters themselves, allowing them to operate independently or support weak grids. This control mode continuously adjusts the converter's output to maintain a stable voltage and frequency by responding to real-time variations in power demand and supply. Through outer control loop, the converter manages real and reactive power, ensuring the system operates within desired limits, while inner control loop fine-tune voltage and current dynamics to maintain performance under changing conditions. Grid-forming control mode is crucial for enhancing the stability and resilience of power systems, especially in grids with a high integration of renewable energy sources or in islanded microgrids \cite{b7}.

The system's dynamic behavior, focusing on the outer and inner control loops for grid stability is described in Eqs. \eqref{8} and \eqref{9}. These dynamics, consistent with the LCL filter, power grid, and average converter model, were used to generate data for evaluating the SINDy and DSR frameworks in directly discovering the system's underlying dynamics from data.

\begin{align}
\textnormal{Outer control}
&\begin{cases}
\frac{1}{\omega_{b}} \frac{d}{dt}{\theta^\textnormal{oc}} &= \omega^\textnormal{oc} - \omega_s\\
\frac{1}{\omega_{z}} \frac{d}{dt}{p}_{m}&= v_{r}^{\textnormal{filt}}i_{r}^{\textnormal{filt}} + v_{i}^{\textnormal{filt}}i_{i}^{\textnormal{filt}} - p_{m} \\
\frac{1}{\omega_{f}} \frac{d}{dt}{q}_{m}&= -v_{r}^{\textnormal{filt}}i_{i}^{\textnormal{filt}} + v_{i}^{\textnormal{filt}}i_{r}^{\textnormal{filt}} - q_{m} \\
\omega^\textnormal{oc}&=\omega^\textnormal{ref} + k_p(p^\textnormal{ref}-p_{m})\\
v^\textnormal{oc}&= v^\textnormal{ref} + 
k_q(q^\textnormal{ref}-q_{m})
\label{8}
\end{cases}
\end{align}
\begin{align}
\textnormal{Inner control}
&\begin{cases}
\frac{d}{dt}{\xi}_{d}&= v_{d}^{\textnormal{vi,ref}} - v_{d}^{\textnormal{filt}} \\
\frac{d}{dt}{\xi}_{q}&= v_{q}^{\textnormal{vi,ref}} - v_{q}^{\textnormal{filt}} \\
\frac{d}{dt}{\gamma}_{d}&= i_{d,}^{\textnormal{cv,ref}} - i_{d}^{\textnormal{cv}} \\
\frac{d}{dt}{\gamma}_{q}&= i_{q}^{\textnormal{cv,ref}} - i_{q}^{\textnormal{cv}} \\
\frac{1}{\omega_\textnormal{ad}}\frac{d}{dt}{\phi}_{d}&= v_{d}^{\textnormal{filt}} - {\phi}_{d} \\
\frac{1}{\omega_\textnormal{ad}}\frac{d}{dt}{\phi}_{q}&= v_{q}^{\textnormal{filt}} - {\phi}_{q} \\
v_{d}^{\textnormal{vi},\textnormal{ref}}&= v^\textnormal{oc} - r_vi_{d}^{\textnormal{filt}} + \omega^\textnormal{oc} l_v i_{q}^{\textnormal{filt}} \\
v_{q}^{\textnormal{vi},\textnormal{ref}}&= -r_vi_{q}^{\textnormal{filt}} - \omega^\textnormal{oc} l_v i_{d}^{\textnormal{filt}} \\
i_{d}^{cv, \textnormal{ref}}&= k_{p}^{v} \left(v_{d}^{\textnormal{vi,ref}} - v_{d}^{\textnormal{filt}}\right) + k_{i}^{v}\xi_{d}  \\&-  c_f\omega^{\textnormal{oc}} v_{q}^{\textnormal{filt}} + k_{\textnormal{ffi}}i_{d}^{\textnormal{filt}}  \\
i_{q}^{\textnormal{cv,ref}}&=  k_{p}^{v} \left(v_{q}^{\textnormal{vi,ref}} - v_{q}^{\textnormal{filt}}\right) + k_{i}^{v}\xi_{q} \\&+  c_f\omega^{\textnormal{oc}} v_{d}^{\textnormal{filt}} + k_{\textnormal{ffi}}i_{q}^{\textnormal{filt}}\\
v_{d}^{\textnormal{cv,ref}}&= k_{p}^{c} \left(i_{d}^{\textnormal{cv,ref}} - i_{d}^{\textnormal{cv}}\right) - \omega^{\textnormal{oc}}\ell_{f}i_{q}^{\textnormal{cv}} \\& + k_{i}^{c}\gamma_{d} + k_{\textnormal{ffv}}v_{d}^{\textnormal{filt}} - k_{\textnormal{ad}}(v_{d}^{\textnormal{filt}}-{\phi}_{d})\\
v_{q}^{\textnormal{cv,ref}}&= k_{p}^{c} \left(i_{q}^{\textnormal{cv,ref}} - i_{q}^{\textnormal{cv}}\right)  + \omega^{\textnormal{oc}}\ell_{f}i_{d}^{\textnormal{cv}} \\& + k_{i}^{c}\gamma_{q} + k_{\textnormal{ffv}}v_{q}^{\textnormal{filt}} - k_{\textnormal{ad}}(v_{q}^{\textnormal{filt}}-{\phi}_{q})
\label{9}
\end{cases} 
\end{align}

\section{Sparse Identification of Nonlinear Dynamics (SINDy)}
SINDy aims to identify governing equations for a system's dynamics that are both interpretable and sparse. Mathematically, it involves identifying a sparse set of nonlinear differential equations for the system's state evolution with control inputs \cite{b8}. 

Consider a dynamical system with n states, and m control inputs denoted as \( \mathbf{x}(t) =[x_1(t), x_2(t), \dots, x_n(t)]^\top \in \mathbb{R}^n \) and \( \mathbf{u}(t) = [u_1(t), u_2(t), \dots, u_m(t)]^\top \in \mathbb{R}^m \). The goal is to identify \( f(\mathbf{x}, \mathbf{u}) \) in a symbolic form while ensuring the solution is sparse. A large set of candidate functions \( \Theta(\mathbf{x}, \mathbf{u}) \) should be constructed, which includes potential nonlinear terms of \( \mathbf{x}(t) \) and \( \mathbf{u}(t) \). This library of functions may consist of monomials, polynomials, trigonometric functions, etc \cite{b4}.

\begin{figure*}
    \centering
    \includegraphics[width=1\linewidth]{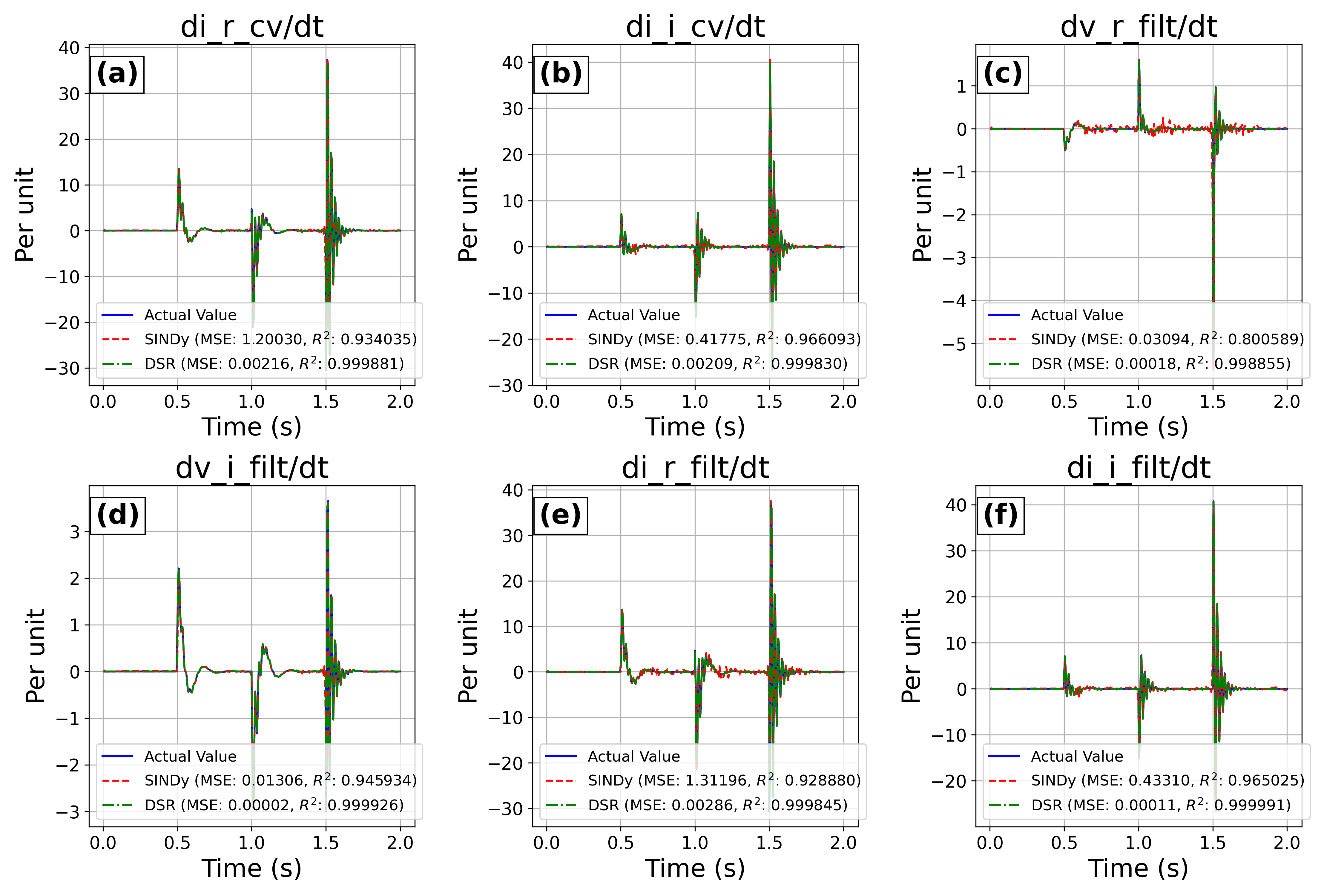}
    \caption{Identification of dynamic response of the LCL filter in a converter infinite bus system operating under grid-forming control mode under three sequential disturbances, using SINDy and DSR methodologies. The disturbances were introduced at different times: a change in active power reference $p^{\text{ref}}$ to 0.7 p.u. at $t = 0.5$ s, a change in reactive power reference $q^{\text{ref}}$ to 0.2 p.u. at $t = 1.0$ s, and a change in voltage reference $v^{\text{ref}}$ to 0.9 p.u. at $t = 1.5$ s. The subplots illustrate the model’s ability to capture the dynamics of various state derivatives under these disturbances; (a)$\frac{d}{dt}{i}_{r}^{\textnormal{cv}}$, (b)$\frac{d}{dt}{i}_{i}^{\textnormal{cv}}$, (c)$\frac{d}{dt}{v}_{r}^{\textnormal{filt}}$, (d)$\frac{d}{dt}{v}_{i}^{\textnormal{filt}}$, (e)$\frac{d}{dt}{i}_{r}^{\textnormal{filt}}$, and (f)$\frac{d}{dt}{i}_{i}^{\textnormal{filt}}$.}
    \label{filter}
\end{figure*}

The system dynamics can then be approximated as a linear combination of these candidate functions:
\begin{equation}
    \dot{\mathbf{x}}(t) \approx \Theta(\mathbf{x}, \mathbf{u}) \xi,
\end{equation}
where \( \xi \) is the vector of coefficients that determines which terms from the library are used to describe the dynamics. A sparse regression problem involves minimizing the least-squares error between actual and predicted values. To ensure the resulting model is sparse, an \( \ell_1 \)-norm regularization term is incorporated into the optimization problem \cite{b8}.

\[
\min_{\xi} \|\dot{\mathbf{x}}(t) - \Theta(\mathbf{x}, \mathbf{u}) \xi\|_2^2 + \lambda \|\xi\|_1
\]

\section{Deep Symbolic Regression (DSR)}
DSR aims to discover underlying mathematical expressions from data by combining symbolic manipulation with the power of deep learning. This hybrid approach allows DSR to search for interpretable mathematical models that fit data while overcoming the limitations of traditional symbolic regression methods, like genetic programming \cite{b5}. 

DSR involves multiple steps to reveal the dynamics of a system, starting with the generation of candidate expressions. It uses a pre-order traversal of a tree structure to represent each expression as a sequence of tokens \cite{b5}. Once the tokens are prepared, an autoregressive recurrent neural network generates mathematical expressions sequentially, token by token. At each step, the network produces a probability distribution over possible tokens, conditioned on previous selections \cite{b5}.
For each token in the expression, the recurrent neural network emits a probability distribution over the set of possible tokens. A token is sampled, and the process repeats autoregressively until the complete expression is generated \cite{b5}. 

When an expression is created, DSR employs nonlinear optimization to fine-tune any constants, ensuring that numerical parameters align closely with data for improved accuracy \cite{b11}. Finally, DSR employs the reinforcement learning framework to train the recurrent neural network, where rewards are based on the expression’s fit to the data. To focus on optimal performance, DSR uses a risk-seeking policy gradient that prioritizes maximizing the best-performing expressions over average outcomes \cite{b11}.

\section{Numerical Results}
In this study, synthetic data were generated in MATLAB using the dynamics associated with grid-forming control mode to evaluate the performance of the SINDy \cite{b4} and DSR \cite{b5} methods in accurately identifying system dynamics. During data collection, three distinct disturbances were introduced at specific times: a change in \( p^{\text{ref}} \) to 0.7 p.u. at \( t = 0.5 \, \text{s} \), a change in \( q^{\text{ref}} \) to 0.2 p.u. at \( t = 1 \, \text{s} \), and a change in \( v^{\text{ref}} \) to 0.9 p.u. at \( t = 1.5 \, \text{s} \). Both SINDy \cite{b8,b9,b10} and DSR \cite{b11} methodologies were then applied to identify the system model based on the generated data. The system parameters used in this study, given in per unit, are as follows: \( X = 0.0020625 \), \( \ell_f = 0.009 \), \( r_f = 0.016 \), \( c_f = 2.5 \), \( \ell_g = 0.002 \), and \( r_g = 0.003 \). 

In practical scenarios, the unavailability of certain states necessitates the use of measurable states for system dynamics identification. To align with real-world applicability, the following states were selected: ${i}_{r}^{\textnormal{cv}}$, ${i}_{i}^{\textnormal{cv}}$, ${v}_{r}^{\textnormal{filt}}$, ${v}_{i}^{\textnormal{filt}}$, ${i}_{r}^{\textnormal{filt}}$, and ${i}_{i}^{\textnormal{filt}}$ (LCL filter); $\theta^\textnormal{oc}$, $p_m$, and $q_m$ (Outer control). This selection ensures that the results are both experimentally viable and relevant for system control and monitoring.

\subsection{LCL filter}
The numerical results pertaining to the LCL filter of the grid-forming control mode is depicted in Figs. \ref{filter}.a, \ref{filter}.b, \ref{filter}.c, \ref{filter}.d, \ref{filter}.e, and \ref{filter}.f. The system response shows significant oscillations following each disturbance event in Fig. \ref{filter}.a. SINDy has a mean square error of 1.2 and $R^2$ score of 0.93, indicating moderate accuracy, while DSR performs better with a mean square error of 0.002 and $R^2$ score of 0.99. The response remains oscillatory with reduced amplitude in Fig. \ref{filter}.b. SINDy captures the trend with a mean square error of 0.4 and $R^2$ score of 0.96, while DSR achieves higher precision with a mean square error of 0.002 and $R^2$ score of 0.99.
    
Small oscillations are observed with minimal deviations after each event in Fig. \ref{filter}.c. SINDy’s mean square error and $R^2$ score (0.003, 0.98) indicate good accuracy, yet DSR achieves near-perfect accuracy (mean square error of 0.00018 and $R^2$ score of 0.99). After each disturbance, a gradual oscillatory decay occurs in Fig. \ref{filter}.d. SINDy has a mean square error of 0.013 and $R^2$ score of 0.99, while DSR provides near-exact predictions with mean square error of 0.00002 and $R^2$ score of 0.99. 

The dynamic oscillations have a greater amplitude in Fig. \ref{filter}.e, and SINDy captures them with a mean square error of 1.3 and $R^2$ score of 0.92. DSR remains accurate, with a mean square error of 0.002 and $R^2$ score of 0.99. However, smaller oscillations follow each disturbance in Fig. \ref{filter}.f. SINDy achieves a mean square error of 0.4 and $R^2$ score of 0.96, while DSR achieves a mean square error of 0.00011 and $R^2$ score of 0.99.

\subsection{Outer control}
\begin{figure*}
    \centering
    \includegraphics[width=1\linewidth]{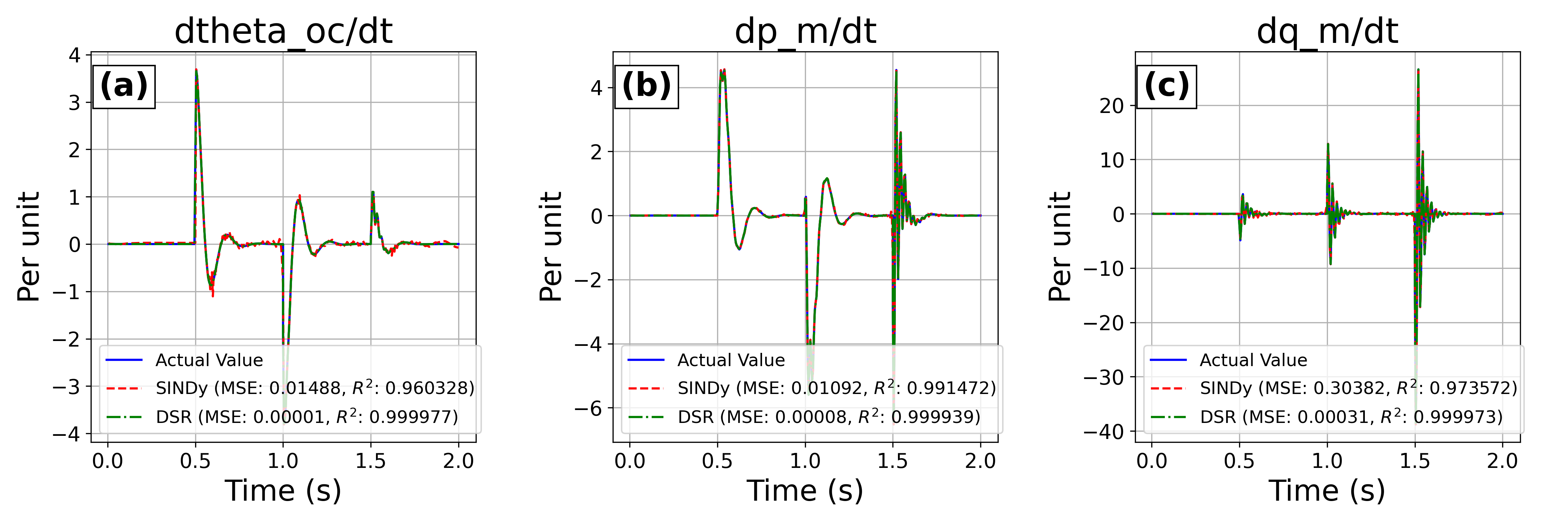}
    \caption{Dynamic response characterization of the outer control loop in a converter infinite bus system operating under grid-forming control mode with three sequential disturbances, utilizing SINDy and DSR methodologies. The disturbances were introduced at different times: a change in active power reference $p^{\text{ref}}$ to 0.7 p.u. at $t = 0.5$ s, a change in reactive power reference $q^{\text{ref}}$ to 0.2 p.u. at $t = 1.0$ s, and a change in voltage reference $v^{\text{ref}}$ to 0.9 p.u. at $t = 1.5$ s. The subplots illustrate the model’s ability to capture the dynamics of various state derivatives under these disturbances; (a)$\frac{d}{dt}{\theta}^{\textnormal{oc}}$, (b)$\frac{d}{dt}{p}_{m}$, and (c)$\frac{d}{dt}{q}_{m}$.}
    \label{outer}
\end{figure*}

The discovered model for outer control loop is presented in Figs. \ref{outer}.a, \ref{outer}.b, and \ref{outer}.c. Small amplitude oscillations occur with a high degree of predictability in Fig. \ref{outer}.a. SINDy’s mean square error is 0.01 with $R^2$ score of 0.96, while DSR performs almost perfectly (mean square error of 0.00001 and $R^2$ score of 0.99). There is a rapid oscillatory decay after each disturbance in Fig. \ref{outer}.b. SINDy achieves good accuracy (mean square error of 0.01 and $R^2$ score of 0.99), as well as DSR provides almost the same results with a mean square error of 0.00008 and $R^2$ score of 0.99. Larger oscillations occur in Fig. \ref{outer}.c, captured moderately well by SINDy (mean square error of 0.3, $R^2$ score of 0.97), with DSR providing closer alignment with mean square error of 0.00031, and $R^2$ score of 0.99.

\section{Discussion}

This study highlights distinct advantages and trade-offs between SINDy and DSR in modeling the dynamics of grid-forming converters, particularly under disturbances. DSR consistently achieves more accurate model, reflected in near-perfect $R^2$ scores ($\geq 0.99$) and minimal mean square error, making it highly effective at capturing complex and nonlinear dynamics such as high-frequency oscillations and rapid transitions. This capability is particularly relevant for grid-forming converters, which often operate under challenging conditions with intricate interdependencies. However, the computational demands of DSR, which are approximately 11 times higher than those of SINDy, limit its practicality for real-time applications. 

In contrast, SINDy offers a balanced trade-off between computational efficiency and accuracy. Its sparse models are not only faster to compute but also more interpretable, a significant advantage for real-time control and system monitoring. This makes SINDy an attractive choice for applications where computational resources are constrained, and quick decision-making is essential. Despite its lower computational cost, SINDy demonstrates robust performance in capturing system dynamics during disturbances, though it occasionally shows moderate deviations in modeling rapid transitions compared to DSR.

The findings also reveal the complementary nature of these methods. While DSR excels in scenarios requiring precise representation of complex dynamics, such as weak or islanded grids with high renewable penetration, SINDy provides a more resource-efficient solution for applications with less stringent accuracy demands. Both methods performed robustly under the tested disturbances, with DSR offering a finer resolution in modeling minute oscillations and decay trends, which could enhance predictive capabilities in preventing system instabilities.

\section{conclusions}
This paper investigates the use of SINDy and DSR for identifying dynamic models of grid-connected converters, focusing on grid-forming control mode. The findings show that while SINDy can generate compact models, DSR offers more accuracy and interpretability in capturing the nonlinear dynamics of converter-based systems. DSR’s strength lies in uncovering system behaviors, making it particularly effective for complex control systems in modern power grids. Future research should develop hybrid frameworks merging SINDy’s efficiency with DSR’s robustness to balance accuracy and computational cost, enabling scalable solutions for complex grid operations with high renewable energy penetration.

\end{document}